\documentclass[preprint,aps] {revtex4}

\usepackage{graphicx}
\usepackage{dcolumn}
\usepackage{amsmath}
\usepackage{bm}
\usepackage{lscape}

\begin{document}

\title {Slow motions detection in polybutadiene through novel analyses of MSE refocusing efficiency and spin-lattice relaxation
}
\author{S. Sturniolo$^{1}$, M. Pieruccini$^{2,\ddagger}$, M. Corti$^{1}$ and A. Rigamonti$^{1}$}
\affiliation{$^1$ Dipartimento di Fisica ``A. Volta'', Universit$\grave{\rm{a}}$ di Pavia, v. Bassi 6, 27100 Pavia, Italy}
\affiliation{$^2$ CNR, Istituto Nanoscienze, v. Campi 213/A, 41125 Modena, Italy}
\author{$^\ddagger$ corresponding author: tel.: +39 059 2055654; fax: +39 059 2055651; e-mail: marco.pieruccini@nano.cnr.it}


\begin{abstract}
Novel methods to analyze NMR signals dominated by dipolar interaction are applied to the study of slow relaxation motions in polybutadiene approaching its glass transition temperature. The analysis is based on a recently developed model where the time dependence in an ensemble of dipolar interacting spin pairs is described without resorting to the Anderson-Weiss approximation. The ability to catch relevant features of the $\alpha$ relaxation process is emphasized. In particular, it is shown that the temperature profile of the Magic Sandwich Echo efficiency carries information on the frequency profile of the $\alpha$-process. The analysis is corroborated by  the temperature dependence of the spin-lattice relaxation time.
\end{abstract}

\includegraphics[width=8 cm, angle=0]{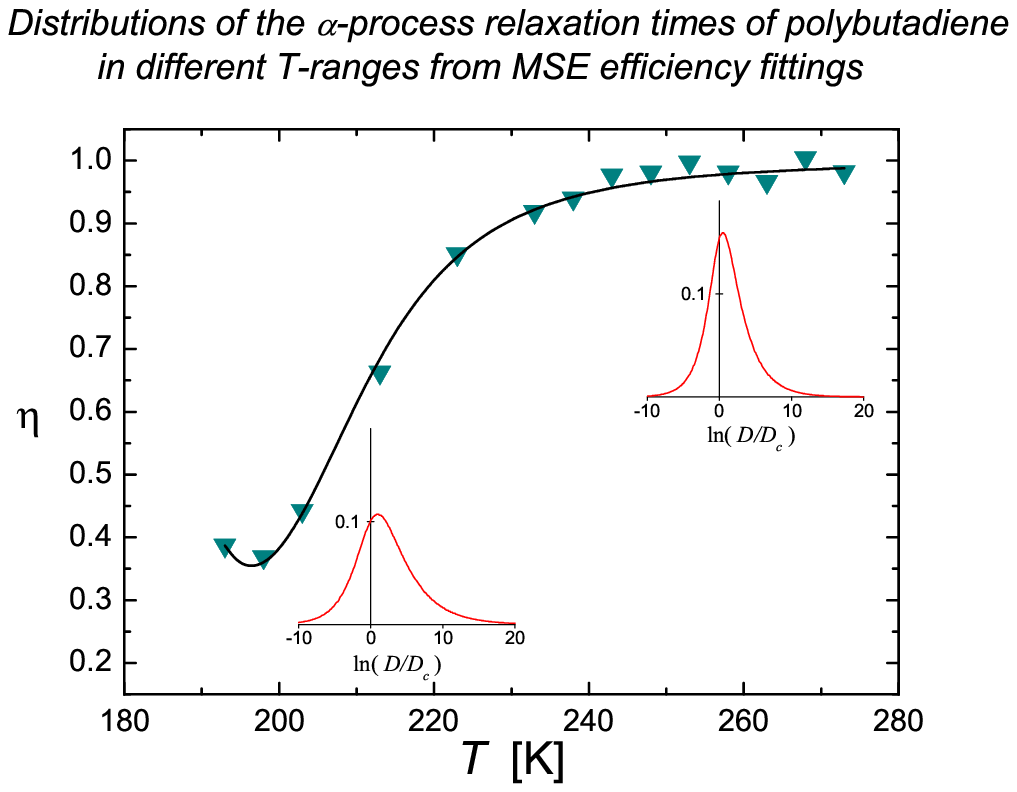}

\maketitle

\section{Introduction}\label{intro}
Polymers are a class of materials which, although very common and thoroughly studied under a variety of aspects, e.g. for industrial applications, still represent suitable model systems for the investigation of some basic processes in condensed matter physics. Amorphous polymers in particular (or the amorphous component of the semicrystalline ones), manifest complex molecular relaxation dynamics covering a wide spectrum of time scales \cite{degenn, doi_edw} which play a role in the kinetic phenomenon known as ``glass transition''. This process entails a dramatic change in macroscopic mechanical properties taking place in a small temperature range around the critical glass transition temperature $T_g$. At microscopic level, the transition is governed by heterogeneous, cooperative, molecular relaxation processes with an unusually strong temperature dependence, often referred to as $\alpha$-process, as opposed to the $\beta$ process, more local in nature, characterized by a somewhat standard Arrhenius-like behavior. These phenomena are not exclusive to polymers, but in the latter the complex underlying molecular structure poses further difficulties for their description if compared to other glass-forming systems \cite{rel_proc_superc}.

In this paper we use a novel approach to analyze H$^1$ NMR data in poly(butadiene) (PB), at the aim of extracting dynamical information on slow cooperative relaxation processes. The choice of PB is motivated by its amorphous nature, its convenient glass transition range and the existence of a wealth of literature data ranging from simulations to NMR and broadband dielectric spectroscopy \cite{gl_diel, PB_decoup, PB_diel_data, PB_liq_melt, PB_liq_melt_2}.

In particular, the analysis of our experimental data is performed by means of a recently developed model \cite{pier_stu_FID} for the magic sandwich echo (MSE) refocusing efficiency, which is known to be affected by slow segmental motion \cite{MSE_stur,NBR_publ,demco}. The worked out information is then used to predict the temperature-dependence of the spin-lattice relaxation rate to check for consistency with independent experimental data.

The results are highly promising and suggest that it is possible to combine the analysis of multiple NMR measurements to derive a reliable picture of slow relaxation dynamics around the glass transition temperature.

\section{Experimental}\label{exp}
Poly(butadiene) was supplied by Polymer Standards Service, Mainz, and composed by poly(1,4 butadiene) with a nominal average molecular weight of $\sim100,000\,\mathrm{g/mol}$, corresponding to $\sim2000$ monomeric units per chain, with minimal dispersity. Its glass transition temperature was around $\sim170\,\mathrm{K}$. The sample was transparent and its consistency at room temperature was of an extremely viscous liquid. Except during experiments, it was always stored at a temperature of $\sim 276$ K; when measurements had to be performed at a distance of days, fresh samples were always used. At relatively high temperatures, extremely sharp, liquid-like lineshapes were observed (corresponding to the long decay times that can be seen in Fig.~2 below). All this guarantees that any crystalline fraction, if present at all, can only appear in traces.

The experiments were carried out using a TecMag ``Apollo'' DoubleResonance Spectrometer, in the working range of 5-450 MHz and a minimum digitization time resolution of 300 ns, and a Bruker BM-10 variable field electromagnet. The measurement chamber was an Oxford CF1200 cryostat able to operate in the temperature range between $4$ and $370\,\mathrm{K}$. Systematic measurements were performed at three different values of the static magnetic field, respectively around 0.5 T, 1 T and 1.5 T. The intensity of the RF pulse used was 30 G ($\pi/2$ pulse duration 2 $\mu$s).

Since the deadtime of the receiver is almost $5\,\mu$s, simple acquisition of the free induction decay (FID) can fail when the decay of the signal is very fast and a significant part of it is lost. To avoid this problem, the FID signal has been refocused using the Magic Sandwich Echo sequence \cite{mse_seminal}. It has been recently shown \cite{NBR_publ} how the MSE refocused FID mantains the same shape as the original one, even when it is scaled down due to molecular motions. The MSE sequence used was the ``non-ideal'' version with a train of $\pi/2$ pulses replacing the long bursts described in the seminal paper \cite{mse_seminal}, at the purpose to avoid problems due to instrumental phase switching times between different pulses. This sequence is substantially equivalent from a mathematical point of view to the original one. A phase switching time of 3 $\mu$s was used (doubled during
the groups of four pulses along the X axis constituting the core of the sequence), and the total length of the MSE sequence was of 96 $\mu$s.

Our analysis was performed with a self-developed software coded in C++. Part of the fittings was performed with the open source program EDDIE (Exact Dipole-Dipole Interaction Estimator which has been released to the public) and that is described in detail in Appendix \ref{app_eddie}.

\section{Theoretical framework}\label{the}

\subsection{Motivation}

At the core of the present paper is a novel aproach to describe the FID signal when the dipolar interaction between protons dominates the Hamiltonian and molecular motion is present.

In order to treat the effect of motion on the FID in the case of polymers, reference is usually done to the chainlike structure of these molecules, and pre-averaging over fast segmental and $\beta$ motion is assumed in deriving analytical expressions for the discussion of experimental data \cite{demco, brer}.

In the attempt to investigate segmental dynamics when $T_g$ is approached from above, however, only the $\beta$ motion can be considered effective in the pre-averaging. Moreover, the manifestation of the cooperative nature of the $\alpha$-relaxation is related to the emergence of constraints which progressively quench the long wavelength components of the chain's collective conformational fluctuations. As a further issue, cooperativity and the glass transition are not exclusive of polymeric systems; thus, referring to a scheme which is more ``local'', in the sense that it is to some extent untied to a chain topology, would be desirable.

For these reasons an expression for the transverse relaxation function $\overline{G}(t)$ has been previously derived \cite{pier_stu_FID}  in the assumption that the system could be represented by an ensemble of spin pairs at a fixed distance with random orientations in space,  uniformly distributed over all the solid angle. Different pairs were assumed independent. Segmental motion was described as an isotropic rotational diffusion, with a diffusion constant $D$.

The prominent mean field character of this model supports its application for a meaningful analysis of the experimental data.  In its crudeness, however, it matches the requirement of locality expressed above and, at the same time, offers the possibility to work out suitable analytical expressions which can be useful to discuss the results, as shown in \cite{NBR_publ}.

In spite of being a rather crude reduction of the complexity characterizing the relaxation processes in polymers \cite{rel_proc_geom}, this model has been substantially adopted to analyze data in similar contexts under the Anderson-Weiss approximation, as in \cite{papon}. In this respect, it is important that in the frame of our model the Anderson-Weiss results are approached when the temperature of the system is sufficiently above $T_g$, as it has been shown in \cite{pier_stu_FID}.

\subsection{Transverse relaxation function}

Following the scheme outlined in the previous SubSection, the transverse relaxation function is expressed by the functional integral 

\begin{equation}\label{fid_avg_mob}
	\overline{G}(t)\equiv \Re \int \delta\psi(\tau)\,p[\psi(\tau)]\, e^{i\int_0^t d\tau\Delta\omega [\theta(\tau)]},
\end{equation}
where $p[\psi(\tau)]$ is the probability associated to an angular trajectory $\psi(\tau)$ of a spin pair during the time $\tau$, and the effect of its orientation with respect to the quantizing magnetic field \textbf{B}$_0$ (i.e. the angle $\theta$) is introduced through the term $\Delta\omega\equiv b\,P_2{\cos [\theta(\tau)]}$ related to the dipolar interaction, with $b$ the coupling constant and $P_2$ the second order Legendre polynomial.

The function $\overline{G}(t)$ can be also recast in the form of an integral over the spin pair orientations at the ends of the time interval $[0,t]$:
\begin{equation}
	\overline{G}(t) = \frac{1}{16\pi^4}\Re\int d\psi_t \,d\psi_0 G[\psi_t,\,t;\,\psi_0,\,0] \,,
\end{equation}
where the Green function $G[\psi_t,\,t;\,\psi_0,\,0]$ is connected to the probability that a spin pair, whose orientation is $\psi_0$ at time $t=0$, ends up with an orientation $\psi_t$ after a time $t$. The function $G$ is initially a Dirac $\delta$-function and progressively it broadens. Its evolution is described by the Dyson equation
\begin{equation}\label{4}
	\begin{array}{l}
		G[\psi_t,\,t;\,\psi_0,\,0] = G_0 [\psi_t,\,t;\,\psi_0,\,0] \,+ \\
		\\
		\,\,\,\,\,\,\,\,\,\,\,\,\,\,\,\,\,\,\,\,\,\,\,\,\,\,\,\,\,\,\,\,\,\,\,\,\,\,\,\,\,\,\,\,\,\, i\,\int_0^t d\tau\int d\psi\, G_0 [\psi_t,\,t;\,\psi,\tau]\,\Delta\omega [\theta,\tau]\,G[\psi,\tau;\,\psi_0,\,0] \,,
	\end{array}
\end{equation}
where the Green function $G_0$ relates to the stochastic evolution dynamics of the spin pair orientation angle $\psi(\tau)$.

The present scheme is general and offers the possibility to consider diverse mechanisms for the evolution of $\psi$ provided that $G_0$ is known. Statistically independent motions affecting the orientation dynamics can be introduced by simple superposition.

When gaussian statistics is being considered, the complexity of working out the transverse relaxation function could be reduced at the outset to some extent, because the relation
\begin{equation}
	\left\langle e^x \right\rangle = e^{\frac{1}{2}\left\langle x^2\right\rangle}
\end{equation}
holds for the average of the associated stochastic variables. This is at the core of the Anderson-Weiss aproximation \cite{A_W}.

To proceed further along the path of the Dyson equation, we chose a rotational diffusion process for the spin pair orientation. Of course, the fact that $G_0$ is known in this case is not of secondary importance; however, there are some further advantages in doing so. One is the possibility to find a check in models based on the Anderson-Weiss scheme, the others relate to some simplifications in estimating quantities of interest such as the MSE refocusing efficiency.

The solution of the Dyson equation can be cast in the form of a series:
\begin{equation}\label{G_resid_1}
	\overline{G}(t) = \Re\left\{ R^{-1}\sum^{\infty}_{k=1} \textrm{res}[W, \omega_k]\, e^{-i\omega_k t}\right\} \,,
\end{equation}
where $\omega_k$ is the $k$-th pole and $\textrm{res}[W, \omega_k]$ the corresponding residue of an appropriate kernel function $W\equiv W(\omega,D,b)$;
\begin{equation}\label{G_resid_2}
	R \equiv \sum^{\infty}_{k=1} \textrm{res}[W, \omega_k]
\end{equation}
represents a normalization factor. In the case where $D=0$ all poles are real.

The function $W$ has been expressed as a continuous fraction \cite{pier_stu_FID}; the $n$ poles and corresponding residues of its $n$-th order rational approximation can be used to form the partial sum
\begin{equation}\label{G_n}
	\overline{G}_n \equiv\Re\left\{R_n^{-1}\sum^{n}_{k=1} \textrm{res}[W, \omega_k] \exp(-i\omega_k t)\right\}
\end{equation}
(with $R_n$ given by the corresponding partial sum in eq.~\ref{G_resid_2}), which reproduces exactly the transverse relaxation function $\overline{G}(t)$ up to a certain time $t_n$. On physical grounds, the truncation means that the evolution process of the whole system is described through a representative finite sub-ensemble of spin pairs [up to $t_n$]. For the typical values of the interaction constant in polymers (of order $\sim100$ kHz) and the usual length of a FID acquisition ($\sim250\,\mu$s), a value of $n\simeq\, 20$ is found suited to fit the data with high precision, but very often (see below) a much lower number of poles suffices.

Analysis of literature data \cite{pier_stu_FID} and new experiments \cite{NBR_publ} indicate that, with regards to direct FID fitting, the model provides results in agreement with those obtained within the Anderson-Weiss approximation for medium and high motional frequencies, while it tends to overestimate orientational diffusivities at low temperatures. For values $D <\,10$ kHz in polymeric samples partial effects of multi-spin interactions due to the high density of protons cause a plateau in the measured diffusivity.

Turning back to the transverse relaxation function, note that $\overline{G}(t_1 + t_2) \neq \overline{G}(t_1) \overline{G}(t_2)$; indeed
\begin{equation}\label{Gt1t2}
	\overline{G}_n (t_1 + t_2) = \Re\left\{R_n^{-1}\sum^{n}_{k=1} \textrm{res}[W, \omega_k] e^{-i\omega_k (t_1 + t_2)}\right\}.
\end{equation}
Therefore, while the coefficients $\textrm{res}[W, \omega_k]$ describe the evolution of the sub-ensemble starting from an initial condition ($t=0$) where the Green function is a $\delta$, the coefficients $\textrm{res}[W, \omega_k]\,\exp\{-i\omega_k t_1\}$ describe the evolution for $t\geq t_1$ from an "`initial"' condition (i.e. at $t=t_1$) where the Green function has already broadened to some extent.

\subsection{MSE efficiency}

MSE allows one to refocus an eco of a dipolar dephased FID with excellent fidelity even long after its decay (more than $100\,\mu s$). If the coupling strength in the system remains constant during the whole experiment, then the refocusing will be complete and the ratio $\eta$ between the intensity of refocused FID to that of the original one is unity ($\eta = 1$). This condition may break down due to molecular motions, causing the decrease of the amplitude of the echo. As a function of the extent of molecular motions, $\eta$ is close to one for frequencies that are very low or very high compared to the order of magnitude of the dipolar coupling constant, being drastically reduced when the two frequencies are of the same order of magnitude \cite{MSE_stur}.

An estimate of the MSE refocusing efficiency may  be obtained following the evolution of the representative sub-ensemble when a pulse sequence $\{\tau_+ |\,4\tau_{-} |\,\tau_+\}$ is imposed, such that in the intermediate interval the time is apparently inverted with regards to the evolution hamiltonian (with the exclusion of the diffusion process, of course). After eq.~\ref{Gt1t2} and the related comments, one finds
\begin{equation}\label{eff}
	\eta = \overline{G}_n (6\tau) = \Re\left\{R_n^{-1}\sum^{n}_{k=1} \textrm{res}[W, \omega_k] \,e^{-i\left(2\left.\omega_k\right|_{\tau_+} - 4\left.\omega_k^*\right|_{\tau_-}\right)\tau}\right\} \,,
\end{equation}
where the poles $\left.\omega_k\right|_{\tau_+}$ and $\left.\omega_k^*\right|_{\tau_-}$ are calculated with coupling constants $b$ and $-b/2$ respectively. (Note the resemblance of eq.~\ref{eff} with eq.~\ref{fid_avg_mob}.) This expression will be subsequently used  for the analysis of the temperature profile of the MSE refocusing efficiency $\eta$. Its relation with the solution of the problem derived by setting $\Delta\omega=bP_2[\Theta(t)-3/2\Theta(t-\tau)+3/2\Theta(t-5\tau)]$ in the Dyson equation (with $\Theta$ the unit step function), is not at all trivial and is currently being subject of a detailed study. In the present context we must limit ourselves to propose it, relying on both the physical argument at the basis of its derivation and the satisfactory analysis presented below.

Figure 1 shows the efficiency as a fuction of $D$ obtained from Eq.~\ref{eff} for different numbers of poles and an MSE sequence of $96\,\mu\mathrm{s}$. Two values of the coupling constant have been considered; one of them is close to that appropriate for two protons a distance $1.8\,\textup{\AA}$ apart ($b=194.107\,\mathrm{kHz}$). Note that a significant dependence on the number of poles only shows up for low values of $D$. This has to be taken into account when extending the analysis of the efficiency data towards $T_g$.

\begin{figure}[h]
\includegraphics[width=9 cm, angle=0]{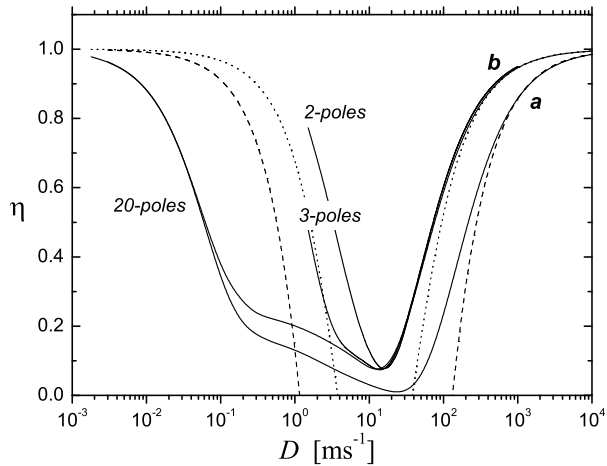}
\caption{\small{
MSE efficiencies calculated from eq.~\ref{eff} for different numbers of poles and for two values of the coupling constant: $b=300\,\mathrm{kHz}$ (curves merging in ``\emph{a}'') and $b=180\,\mathrm{kHz}$ (curves merging in ``\emph{b}''). Dash and dotted lines represent the efficiencies calculated with eq.~\ref{EffAW} with the van Vleck second moment given by $M_2 = b^2/5$.
}} \label{fig1}
\end{figure}

The efficiencies predicted by eq.~\ref{eff} are also compared with the expression reported below, which has been derived in~\cite{demco} within the Anderson-Weiss approximation:
\begin{equation}\label{EffAW}
	\eta = 1 - M_2 \tau_c^2\left[ e^{-6\tau/\tau_c} - 3e^{-5\tau/\tau_c} + \frac{9}{4}e^{-4\tau/\tau_c} + 3e^{-\tau/\tau_c} + \frac{3\tau}{\tau_c} - \frac{13}{4}\right] \,.
\end{equation}
Since $\tau_c$ is in fact the relaxation time of $\left\langle P_2\right\rangle$, the relation $\tau_c=(6D)^{-1}$ holds~\cite{pier_stu_FID}. Equation~\ref{EffAW} is strictly valid for $\eta$ close to unity~\cite{demco}, but has been plotted in the whole $D$ interval to point out how it compares with eq.~\ref{eff} (setting its value to zero wherever negative, of course). From a qualitative point of view the efficiencies calculated with the two above expressions are similar. The extrapolation of eq.~\ref{EffAW} to low $\eta$ values underestimates the efficiency predicted by eq.~\ref{eff} unless $D$ is small enough; then, the behavior is reversed. This compensation may play some role when average efficiencies calculated with eqs.~\ref{eff} and~\ref{EffAW} are fitted to the data, as the results are found similar to some extent (see below).

\subsection{Spin-lattice relaxation}

Spin-lattice relaxation times will be considered for testing the parameters of the motional distribution derived from the analyses of the FIDs and of the MSE efficiency. In particular, the T$_1$ data as a function of  temperature will be compared with those obtained by the equation
\begin{equation}\label{T1_dipolar}
 \frac{1}{\textrm{T}_1} = \frac{9}{8}\frac{\gamma^4\hbar^2}{r^6}\left(\frac{\mu_0}{4\pi}\right)^2\left[\left\langle J^{(1)}(D_c, \omega_L)\right\rangle + \left\langle J^{(2)}(D_c, 2\omega_L)\right\rangle\right]
\end{equation}
where
\begin{equation}\label{J_integr}
  \left\langle J^{(i)}(D_c, \omega_L)\right\rangle \equiv \int_{-\infty}^{\infty}{g(\mathrm{ln}D,D_c)J^{(i)}(D, \omega_L)d\mathrm{ln}D},
\end{equation}
are taken as superpositions of single-$D$ contributions. The shape of the distribution $g(\mathrm{ln}D, D_c)$ is determined by the parameters worked out from the MSE efficiency; $D_c$ is the central relaxation rate (rotational diffusivity in our case) for the distribution. The integral is carried over the logarithm of the frequency, according to a linear distribution of energy barriers \cite{logMOM}. As for the temperature dependence of $D_c$, two different choices have been considered. More details will be given in the following Section, dealing with the experimental data.

\section{Experimental Results}\label{res}

\subsection{FID}\label{res_FID}

The analyses of the FIDs have been performed both on the basis of our model and with the expression below, derived within the Anderson-Weiss approximation~\cite{papon}:
\begin{equation}\label{FID-AW}
	I_{\scriptscriptstyle{\rm{FID}}}(t) = \exp \left[-M_2 \tau_c^2\left(e^{\tau/\tau_c} + \frac{t}{\tau_c} -1\right)\right] \,.
\end{equation}

Following the lines of Ref.~\cite{papon}, the value of $b\simeq 300$~kHz which has been obtained from the $T=173$~K FID fitting (i.e. $\sim300$ kHz using $G_n$ and $\sim275$~kHz using eq.~\ref{FID-AW}) has been taken as a fixed parameter for the analysis of the higher temperature FIDs. The values of the diffusion constant worked out with our expression were found to be practically the same as those obtained with eq.~\ref{FID-AW} for $T \gtrsim 190$~K. For some intermediate temperatures around $T=213$~K, the quality of the fittings was found to degrade slightly, independent of the analytical expression used for the FID. Considering a distribution of $D$ for such cases (i.e. a Gaussian) could hardly improve the fittings a little, and not in all cases. By the way the worked out average $D$ did not significantly differ from that derived using a single-$D$ expression.

Some of the FIDs and their fittings are shown in Fig.~2. As it can be seen, the crossover between a rigid-like dipolar dephasing and a mobile FID decay takes place somewhere around $200\,\mathrm{K}$, above the calorimetric glass transition temperature. This is consistent with the fact that this process is sensitive to a shorter time scale, around $10\,\mu\mathrm{s}$, while the ordinary macroscopic techniques used to assess the glass transition, probe motions on a time scale of seconds.

\begin{figure}[h]
\includegraphics[width=9 cm, angle=0]{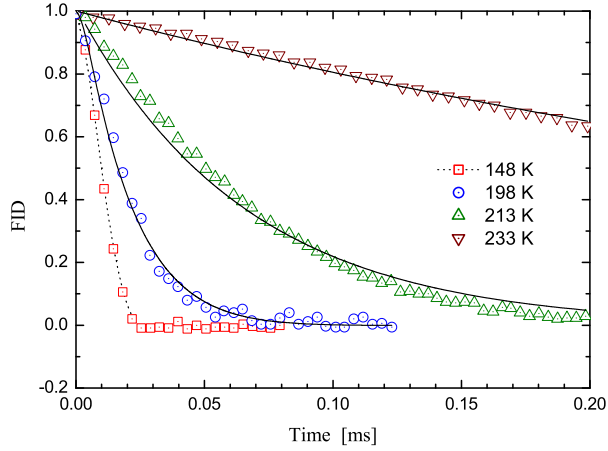}
\caption{\small{
MSE refocused FIDs in PB at different temperatures in the static field of $22\,\mathrm{MHz}$ (symbols) and their respective fittings according eq.~\ref{G_n} (lines) for $n = 20$.
}} \label{fig2}
\end{figure}

The fitted values of the diffusivity $D$ are shown in Fig.~3. A finite plateau with $D$ of the order of 30 ms$^{-1}$ at low temperatures was found, which is likely due to competitive relaxation process and/or multi-spin interactions. Using eq.~\ref{FID-AW} above, however, didn't improve the situation significantly since, e.g., best fit values of $D=$ 7.6, 15 and 17~ms$^{-1}$ were obtained from FIDs at $T=$ 173, 178 and 183~K respectively.

\begin{figure}[h]
\includegraphics[width=10 cm, angle=0]{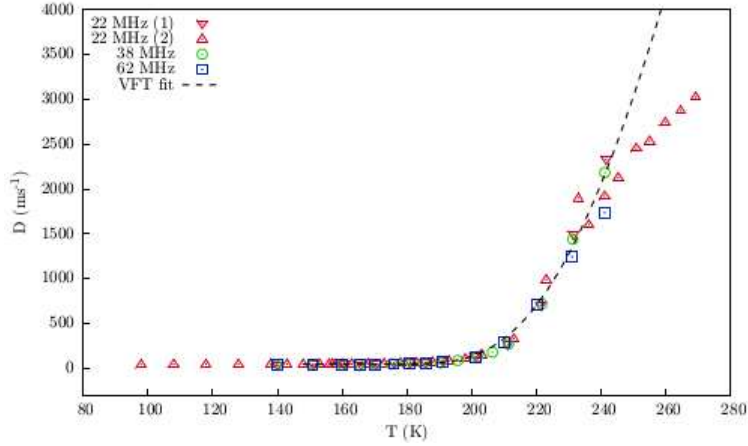}
\caption{\small{
Values of the rotational diffusivity $D$ for different applied static magnetic fields; at 22 MHz there are two different sets, which led to almost identical results. At low temperatures a plateau of $\sim30\,\mathrm{ms^{-1}}$ is found. At high temperature the data bend due to FID sizeably affected by magnetic field inhomogeneities, as expected. The dashed line represents a fitting Vogel-Fulcher-Tammann law.
}} \label{fig3}
\end{figure}

For temperatures above $T_g$ a steepy growth of $D$ is observed, and a maximum limit is reached once the motions reach frequencies so that the dipole-dipole interaction is averaged out and field inhomogeneities dominate the line-width and then the related FID's. No change was found fitting FIDs measured at different values of the static magnetic field in the temperature of interest.

A Vogel-Fulcher-Tammann (VFT) law,
\begin{equation}\label{VFT_form}
 D(T) = \frac{1}{\tau_{\infty}}\exp\left(\frac{-AT_{\scriptscriptstyle{\rm{VFT}}}}{T-T_{\scriptscriptstyle{\rm{VFT}}}}\right) \,,
\end{equation}
has been adjusted to the data and the following values of the best fit parameters have been obtained: $\tau_{\infty}=8.86\cdot10^{-9}\,\mathrm{s}$, $T_{\scriptscriptstyle{\rm{VFT}}} = 148\,\mathrm{K}$ and $A = 2.5$. The resulting expression is plotted as a dashed line in Fig~3.

\subsection{MSE efficiency} \label{res_MSE}

As already mentioned, direct evaluation from the fitting of the FID leads to overestimate  the $D$ values, particularly at low temperatures. Thus, in order to work out reliable information from the analysis of the MSE efficiency, $D$ vs. $T$ data obtained from dielectric spectroscopy \cite{PB_diel_data} have been used, and the temperature dependence of the $\alpha$ relaxation in the $T$-range of interest has been taken into account. (We note that VFT extrapolations and experimental values for the $\beta$-process reported in~\cite{PB_diel_data}, indicate that the latter can be considered much faster than the $\alpha$ relaxation only marginally at $T=273$~K, i.e. the highest temperature value explored in our measurements.)

A first assessment was made by considering a single-$D$ relaxation, with a $D$ vs. $T$ dependence given by the VFT parameters provided by~\cite{PB_diel_data}, namely, $\tau_{\infty} = 4.8\cdot10^{-13}\,\mathrm{s}$, $T_{\scriptscriptstyle{\rm{VFT}}}=142\,\mathrm{K}$ and $A=7.96$.

In Fig.~4 the MSE efficiency measured for PB is reported, and compared with the expressions given by eq.~\ref{eff} calculated for $n=20$ and by eq.~\ref{EffAW} in the hypothesis of single relaxation time.

\begin{figure}[h]
\includegraphics[width=10 cm, angle=0]{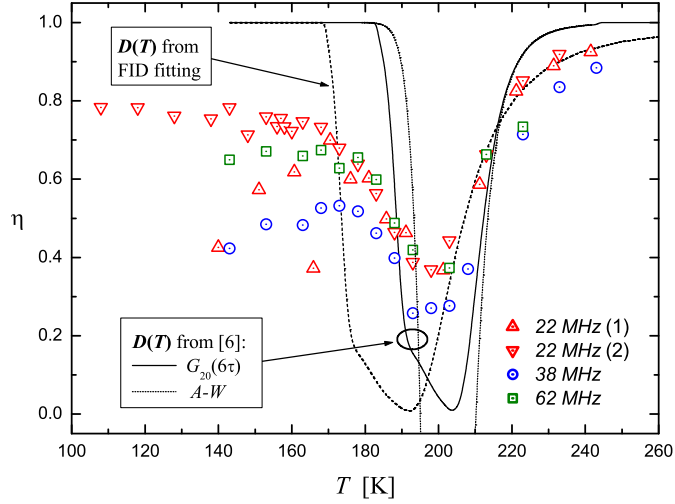}
\caption{\small{
MSE efficiency measured in PB for different values of the static magnetic field (symbols). The solid and the dashed lines are theoretical efficiencies obtained from eq.~\ref{eff} calculated for $n=20$ using the $D_c$ vs. $T$ dependencies from ref.~\cite{PB_diel_data} and from FID fitting respectively; the dotted line has been obtained from eq.~\ref{EffAW} ($b=300$ kHz).
}} \label{fig4}
\end{figure}

As it can be seen, the predictions remarkably match the temperature range where one has the efficiency dip. Better approximations can be obtained considering a distribution of relaxation times. To this aim, for each temperature the efficiency was calculated as an integral over the logarithm of the rotational diffusion constant (cf. the linear dependence of $\overline{G}$ on $p[\psi(\tau)]$ in eq.~\ref{fid_avg_mob}):
\begin{equation}\label{eff_distr}
 \eta(T) = \int_{-\infty}^{\infty}{F_{\alpha}(D, D_{\alpha}(T))\eta(D)}\, d\,\mathrm{ln}D
\end{equation}
where $F_{\alpha}$ is the (normalized) distribution associated to the $\alpha$-process and is taken in the form
\begin{equation}\label{distrHN}
F_{\alpha}(D, D_c) = \frac{1}{\pi}\frac{\left(D_c/D\right)^{ac}\sin(c\theta)}{\left[1+2\left(D_c/D\right)^a \cos\left(\pi a\right)+\left(D_c/D\right)^{2a}\right]^{-c/2}}   ,
\end{equation}
where
\begin{equation}\label{theta_1}
\theta = \mathrm{atan} \left[ \frac{\sin(\pi a)}{\left( D_c/D \right) ^{a}+\cos(\pi a)}\right]
\end{equation}
if the argument of the arctangent is positive and
\begin{equation}\label{theta_2}
\theta = \mathrm{atan} \left[ \frac{\sin(\pi a)}{\left( D_c/D \right) ^{a}+\cos(\pi a)}\right] + \pi
\end{equation}
otherwise; $a$ and $c$ (both positive and not larger than one) are the width and symmetry parameters of the distribution.

The reason for assuming eq.~\ref{distrHN} is that $F_{\alpha}$ ``generates'' the Havriliak-Negami distribution, i.e.
\begin{equation}\label{HN_integral}
\frac{1}{[1 + (i\omega/D_c)^a]^c} = \int_{-\infty}^{\infty}{\frac{1}{1+i\omega/D}}F_{\alpha}(D, D_c)\,d\,\mathrm{ln}(D) \,,
\end{equation}
which is used very often to fit relaxation processes in dielectric spectroscopy. Equations \ref{HN_integral} and \ref{J_integr} share the same structure, with the difference that the single frequency spectral density is replaced here by the dielectric response of a Debye process.

The single-$D$ behavior reported in Fig.~4 suggests that data analysis can only be performed for temperatures approximately above $188$~K if no other mechanism controlling the MSE efficiency is included on top of the one considered. Thus we limit data fitting to the interval $T \geq 193 K$ (except in one case, where also the efficiency at $T=188$~K has been considered) and just for the 22 MHz data set. The solid lines in Fig.~5 refer to the efficiency expressed by eq.~\ref{eff} in the case where the VFT parameters of $D_c(T)$ were given the values of either ref.~\cite{PB_diel_data} (line $a$) or those derived from the analysis of the FIDs (line $b$). A rather bad performance is evident in the latter case.

\begin{figure}[h]
\includegraphics[width=10 cm, angle=0]{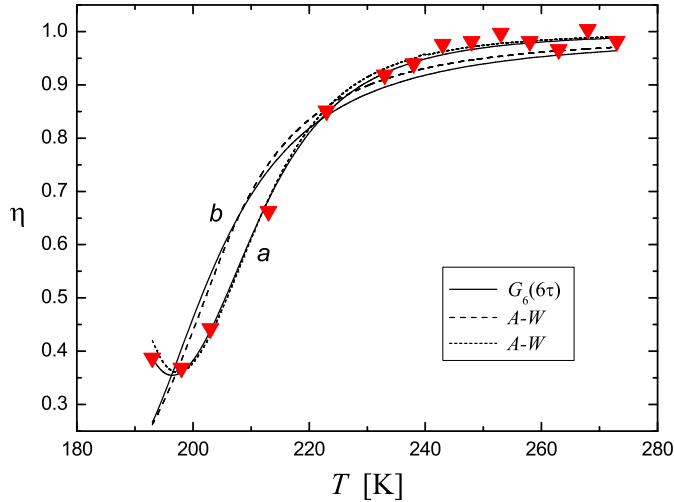}
\caption{\small{
MSE efficiency at 22 MHz. Solid lines fit the data according eq.~\ref{eff}, dashed lines are obtained according to eq.~\ref{EffAW}.
}} \label{fig5}
\end{figure}

Notwithstanding the fact that some integration subinterval is out of the domain of validity of eq.~\ref{EffAW}, the same calculation has been performed within the Anderson-Weiss approximation. The results are shown in Fig.~5 as dashed lines, again derived assuming the two $D_c(T)$ VFT dependencies as above. The striking similarity between the two forms of $\eta (T)$ seems to suggest that the relevant character for a description of the experimental data is a \emph{qualitative} nature of the $\eta$ vs. $D$ dependence. Table~\ref{tab1} reports the best fit values of the Havriliak-Negami parameters obtained so far.

\begin{table}
\begin{center}
\begin{tabular}{| c | c | c | c |}
\hline
				&								&					 &					\\
 ~~~~$D_c(T)$~~~~ & ~~~~$T$-range (K)~~~~ & ~~~~$\overline{G}_6(6\tau)$~~~~ & ~~~~AW~~~~ \\
  			&								&					 &					\\
\hline
				&								&					 &					\\
			  &    193-273    & $a=0.50$ & $a=0.53$ \\
				&								& $c=0.61$ & $c=0.52$ \\
				&								&					 &					\\
		  	&    223-273    & $a=0.56$ & $a=0.56$ \\
  			&								& $c=0.69$ & $c=0.59$ \\
 from	\cite{PB_diel_data}	&								&					 &					\\
	  		&    193-223    & $a=0.49$ &     -    \\
				&								& $c=0.63$ &     -    \\
				&								&					 &					\\
			  &    188-213    & $a=0.45$ &     -    \\
				&(with 10 poles)& $c=0.59$ &     -    \\
				&								&					 &					\\
\hline
				&								&					 &					\\
from FID&    193-273    & $a=0.80$ & $a=0.82$ \\
fitting	&								& $c=0.67$ & $c=0.54$ \\
				&								&					 &					\\
\hline
\end{tabular}
\end{center}
\label{a_c_table}
\caption{Values of the $\alpha$-relaxation  parameters $a$ and $c$ obtained by fitting the MSE efficiency in different temperature ranges. All of them have been derived assuming the $D_c(T)$ dependence provided by ref.~\cite{PB_diel_data}, except in the last line, where $D_c(T)$ is a VFT tracing the $D$ values obtained from fitting the FIDs.}
\label{tab1}
\end{table}

Fittings have been also performed within different temperature subintervals, and the worked out values of the Havriliak-Negami parameters are reported in the table. It is worth noticing that width and asymmetry of the $F_\alpha$ profile increase when the average temperature of the fitting interval decreases. This behavior is  more evident in Fig.~6.

\begin{figure}[h]
\includegraphics[width=10 cm, angle=0]{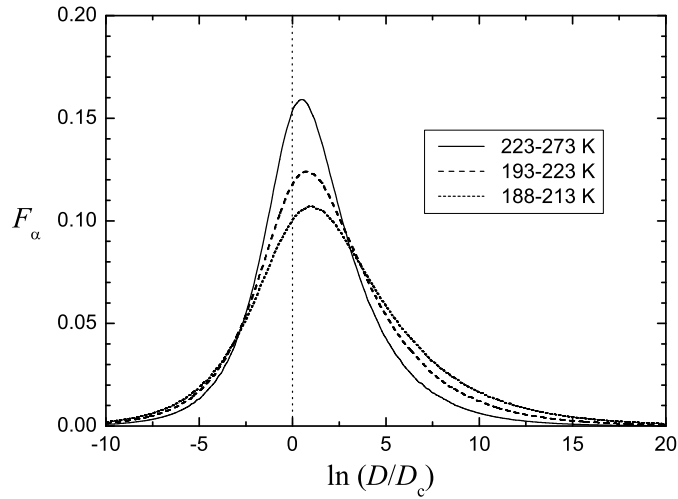}
\caption{\small{
Profiles of the function $F_\alpha$, generating the Havriliak-Negami distribution, for different temperature intervals. On decreasing the average temperature of the fitting interval the profile broadens and becomes less symmetric.
}} \label{fig6}
\end{figure}

With regards to eq.~\ref{EffAW}, instead, the analysis is limited to a comparison between the results worked out on the whole $T$-interval and the high-$T$ region for obvious reasons. As is evident, the changes in the parameters is not as pronounced as in the case where eq.~\ref{eff} is used.

The values of the shape parameters derived in the whole $T$-range can be also compared with those derived from dielectric analysis in ref.~\cite{colmenero}, namely, $a=0.72$ and $c=0.50$.

\subsection{Spin-lattice relaxation}\label{res_T1}

Finally, the spin lattice relaxation time as a function of the temperature has to be discussed. The presence of a broad relaxation distribution underlying the spin lattice process can be qualitatively guessed already from the fact that the maxima in the relaxation rates are inversely proportional to the strength of the field, namely
$(1 / T_1)_{max} \propto \omega_L^{-1}$ (with $\omega_L$ the Larmor frequency) rather than to the square of the inverse $\omega_L$, as expected in the case of single frequency characterizing the dynamics. The analysis was carried out by using the method described in Section \ref{the}, eqs.~\ref{T1_dipolar} and~\ref{J_integr}. With reference to Fig.~7a, for each set of data three curves have been plotted: two of them correspond to the same $D_c(T)$ dependence provided by ref.~\cite{PB_diel_data} but two different $T$-intervals for the efficiency fits; the other one is obtained taking the VFT parameters of Fig.~3. A constant baseline of 3 s$^{-1}$ has been added to all curves for a rough account of all those faster processes that cause relaxation at low temperatures but are not described by our motional distribution. The shape parameters used to draw the lines are those derived from the analysis of the 22 MHz MSE efficiency, for this reason some mismatch can be found with the 62 MHz data. This has been done at the aim to assess how the spin-lattice relaxation profile can be inferred from the analysis of the MSE efficiency in different conditions.

\begin{figure}[h]
\includegraphics[width=10 cm, angle=0]{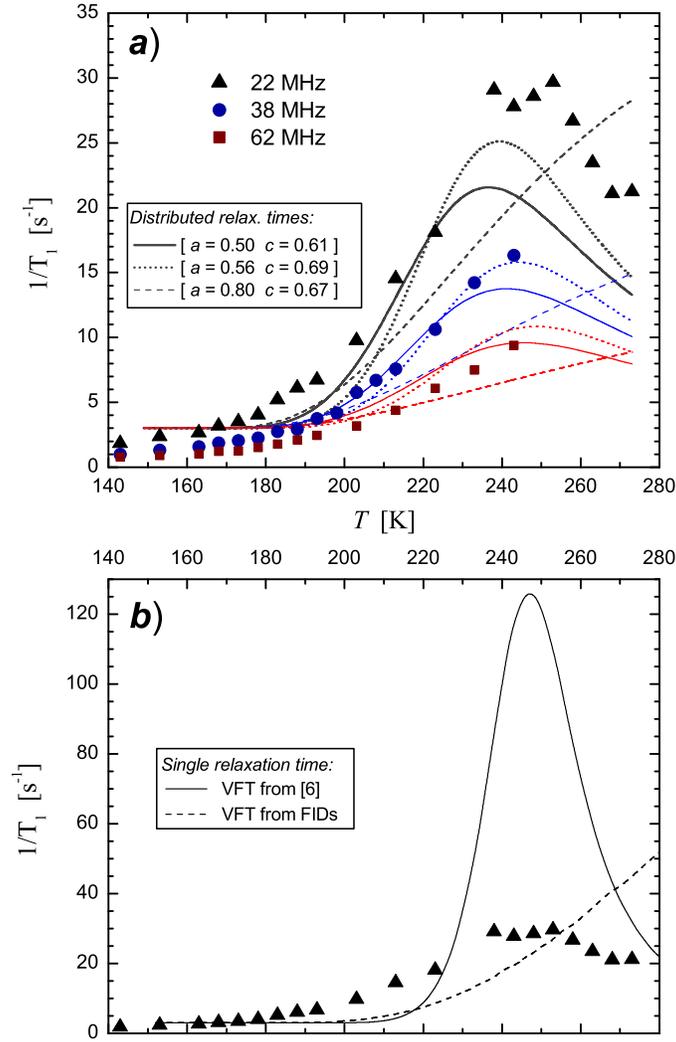}
\caption{\small{
a) Spin-lattice relaxation rates for three values of the static magnetic field. The dashed lines have been obtained from a narrow distribution assuming mean relaxation times described by the VFT function fitting the data in Fig.~3. Solid and dotted lines have been obtained for shape parameters worked out from MSE fittings in the whole-$T$ interval and in the high-$T$ interval respectively, with the VFT parameters of ref.~\cite{PB_diel_data}. b) 22 MHz data compared with the T$_1^{-1}$ vs. $T$ behaviors in the case of a single relaxation time and for the two sets of VFT parameters.
}} \label{fig7}
\end{figure}

Curves corresponding to  narrow $D$ distributions ($a$ close to unity) manifestly fail to catch the corresponding T$_1^{-1}$ maximum. Considering the 22 MHz data (to which indeed the shape parameters refer), it is evident that the maximum is better approached when the high-$T$ interval shape parameters are taken, i.e. for a moderately narrower distribution. On the contrary, the low-$T$ data are better described with a broader relaxation time distribution. This indicates that also in this circumstance, accounting for the appropriate $T$-dependence of the shape parameters would be desirable to improve the fittings, and that providing just their ``mean'' values worked out from such wide $T$-range does not give a detailed analysis of the $\alpha$-relaxation.

Figure 7b reports the T$_1^{-1}$ vs. $T$ dependence ($\omega_L=22$~MHz) obtained considering a single relaxation time ($a=1$), with either the VFT parameters of ref.~\cite{PB_diel_data} or those obtained from the fittings of the FIDs. In the former case a maximum of $\sim 130$~s$^{-1}$ is found at a temperature around 247 K; in the latter, the temperature of the maximum shifts to a value of $\sim 470$~K.

\section{Concluding remarks}

A relevant issue emerging from the analysis carried out so far is that the frequency profile of the $\alpha$ process can be extracted from the $T$-dependence of the MSE refocusing efficiency. Most importantly, the analysis is able to reveal the temperature dependence of the relaxation time distribution characterizing the $\alpha$-process; moreover, the results are consistent with the expected trend. In fact this kind of relaxation can be generally described by means of a stretched exponential $\exp\{-(t/\tau_{\scriptscriptstyle{\rm{KWW}}})^\beta\}$ (the Kohlrausch-William-Watts function), with the exponent $\beta\leq 1$ decreasing on lowering the temperature \cite{phillips}. On the other hand, from the relation $ac\approx\beta$ \cite{colmeneroHN} it is easy to check that our best fit values of the Havriliak-Negami parameters follow the expected temperature behavior.

The adopted model catches the main features of schemes derived within the Anderson-Weiss approximation. In this sense, the good matching found of the results has to be considered a valuable support to our theory. The results reported here, further extend the agreement to the analysis of the MSE refocusing efficiency.

Referring to some specific issues of the present report, we consider first the rather crude estimate of $D$ from the fittings of the FIDs. The results show that at low temperatures our model overestimates this quantity with respect to the case where the Anderson-Weiss based model is used. Apart of this modest discrepancy, both approaches lead to substantially the same VFT parameters for the $D$ vs $T$ dependence, which on the other hand differ quite markedly from those found from dielectric analysis.

The rather good results obtained with dielectric VFT parameters on \emph{both} the T$_1$ and MSE efficiency profiles, indicate an inconsistency in the FID analysis, or at least in its interpretation. We don't want to analyze this aspect in detail in the present context, but note that the \emph{single relaxation time} efficiency ($a=c=1$) accommodates very well among the data when $D(T)$ as obtained from FID's analysis is used (and if the minimum is not approached too closely; see Fig.~8, showing a detail of Fig.~4). Of course this occurrence is not significant with regards to the spin-lattice relaxation profile, as the maximum wouldn't appear in the relevant $T$-range.

\begin{figure}[h]
\includegraphics[width=10 cm, angle=0]{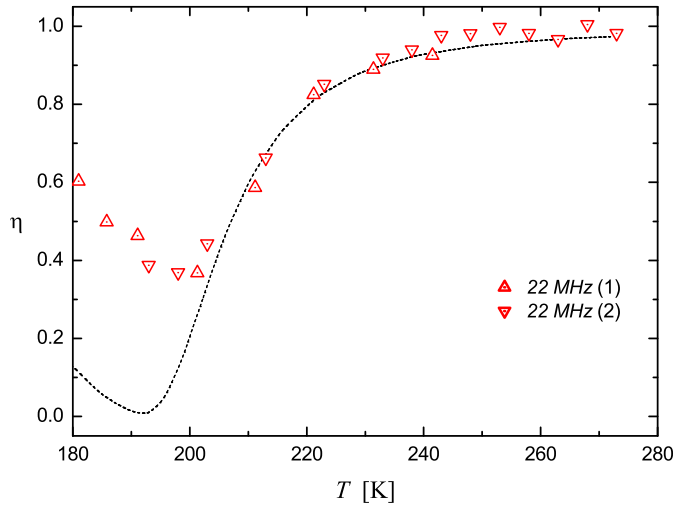}
\caption{\small{
Efficiency data at 22 MHz and single time efficiency from eq.~\ref{eff} with VFT parameters for $D_c(T)$ derived by FID fittings.}} \label{fig8}
\end{figure}

Overall, $D(T)$ is progressively overestimated as $T$ decreases, also reaching a finite plateau below $T_g$. This is motivated by at least a basic assumption of the model, namely, that $\Delta\omega[\theta(\tau)]$ in Eq. \ref{fid_avg_mob} only involves a single spin pair. In reality a given spin interacts also with others at comparable distances. Thus the fluctuations in $\Delta\omega$, which are due to multispin interaction, are ascribed by the model to the motion of just one pair. This circumstance and the fact that at long time an increasing number of spins correlates, might well be at the origin of the problem, becoming ever more important as the motion is slowing down, i.e. as $T \rightarrow T_g$. (See e.g. \cite{correlations} for solids; roughly speaking, the effect of a neighbouring spin starts being significant after a time of the order of the inverse coupling constant.)

On the other hand, each spin pair interaction term in the multispin hamiltonian is virtually ``reversed'' by the pulse sequence, and contributes separately to the formation of the echo (i.e. the effect of the reconstruction on the density matrix factorizes). This means that in this circumstances the response of the system differs very little from that of our simplified model. For this reason the introduction of a ``correct'' $D_c$ vs. $T$ dependence appears to be crucial for a reliable analysis of the refocusing efficiency.

The need to take into account other experimental techniques (such as dielectrics in this case) is not uncommon in polymer physics, and it has been crucial for highlighting the possibilities offered by a proper analysis of the MSE efficiency. From the practical point of view, however, the present state of the art is rather unsatisfactory, since a complete, self contained analysis of the slow motions via H$^1$ NMR would be desirable (e.g. in those cases where dielectric analysis would be difficult, like in polyolefines). In this respect, the present results address the opportunity of extending the model to account for multispin effects and render the direct FID analysis more reliable. As another issue, noting how a correct $T$-dependence of the central relaxation time affects the quality of the fittings overall, in particular with the respect to the correct matching around the minimum of the efficiency (see Figs.~5 and~8), the possibility to derive good VFT parameters from a joint analysis of the efficiency and of the spin-lattice relaxation rate (which is sensitive to the shape parameters of the distribution) seems to be a reasonable target. Work in this direction is underway.

In concluding, the results reported in the present paper are highly promising and indicate the possibility of reliable analyses based on multiple NMR data in order to extract significant insights on the dynamics around the glass transition temperature in polymers.

\section{Acknowledgements}

S.Sturniolo gratefully acknowledges extensive discussions with Prof. K. Saalw\"{a}chter about MSE refocusing efficiency and its use for the analysis of slow motion in H$^1$ NMR experiments.

\clearpage

\appendix

\section{Fitting software}\label{app_eddie}

This Appendix contains technical details about EDDIE, the fitting software developed in order to implement the theoretical approach. EDDIE stands for Exact Dipole-Dipole Interaction Estimator. We will provide all the necessary details as well as a short user manual. The program has been released to the public at the address https://sites.google.com/site/eddienmr/home.\newline

\subsection{Technical details}\label{appD_EDDIE_tech}

The purpose of the EDDIE software is to simulate, given the parameters $b$ and $D$, a FID signal using the function $\overline{G}(t)$ of Section \ref{the} (reported explicitly in \cite{pier_stu_FID}) and fitting raw experimental data with it. We can summarize the steps required to generate a FID signal as follows:

\begin{itemize}
 \item choose a value for the parameters $b$ and $D$, as well as the number of poles $n$, which will determine the precision of the final result;
\item calculate the kernel of the anti-fourier transform;
\item find the roots in $\omega$ for the denominator of the kernel (an equation of $n$-th degree), thus identifying the poles;
\item calculate the residues of the kernel in each of the poles;
\item calculate the FID by performing a proper sum over the residues, with the formula of eq. \ref{G_n}.
\end{itemize}

Fitting the experimental data requires an additional step. In fact, since the function is not analytical, our best option is to use a simplex method to calculate the FID in fixed points of the $b-D$ plane and then refine our search using the residual sum of squares as a parameter to minimize. The procedure has to be iterated many times, as the fitting simplex moves in the function's domain, and becomes the most time-consuming step of the entire process. However this is not too demanding and the program, running on a common laptop, is able to perform a 1000 step simplex fitting in less than a minute.

The program accepts $b$, $D$ and $n$ as user input; in lack of an input, it has default values for them. The following calculation of the kernel requires operations between complex polynomials. EDDIE was written in C++.  In order to simplify portability, dependencies from external libraries were avoided. Therefore, the software comes with its own \textit{complex} and \textit{polynomial} classes with overloaded operators.

The solution of the denominator of the kernel is carried out by using a Jenkins-Traub algorithm. In this case we are using with permission a version for complex roots written by Henrik Vestermark and publicly available on his site \cite{hvks}. From our tests, and in our specific case, the algorithm seems to give reliable results only for polynomials up to the 35th degree. We tried improving that by changing algorithm or making use of high precision libraries, without success. This does not constitute a major problem as a number of poles of 20 or so gives excellent precision for most practical applications. After the poles are found (a \textit{FID} class has been written that keeps the solutions in memory after finding them as part of its initialization), the procedure is rather straightforward, as by inserting the desired time $t$ it is automatically possible to carry out the sum in eq. \ref{G_n} and thus find the FID.

The fitting procedure, as mentioned before, makes use of a simplex algorithm. Since it is important to carry out the procedure in a limited domain of possible values and the function to minimize is rather difficult to handle, as it has many local minima which can cause wrong fitting, a special algorithm developed to be a ``constrained, global and bounded Nelder-Mead method'' by Luersen et al. was used \cite{GBNM}. This algorithm works like a regular Nelder-Mead simplex optimization procedure, with a few differences:

\begin{itemize}
\item the search goes on for a predefined number of steps rather than waiting for a condition to be satisfied;
\item the simplex is constrained to stay inside a fixed domain - the search is interrupted and restarted if it either finds a local minimum or if it gets stuck or deformed by the boundaries;
\item in order to improve the probability of finding a global minimum, every time a local minimum is found, a bias function is calculated in order to make it less likely that the search restarts from the vicinity of that point.
\end{itemize}

\subsection{User manual}\label{appD_EDDIE_UM}

EDDIE can be operated in two ways: by console or with an input file. The two methods are very similar, with only a few differences. Besides fitting and simulating a FID it is possible to verify or change the values of the internal variables of the program, which control parameters like the boundaries of the fitting, the number of poles used, etc. Each variable is initialized to a default value when the program is started: console commands allow to interact with them. The program is run from the system console by simply typing its name and hitting return:


user:~\$ eddie

\noindent
At this point, the internal console of EDDIE will show up. In this context, the general syntax for any command is:


  > [COMMAND] [ARGUMENTS]

\noindent
Commands and arguments are case-sensitive. This is a list of possible commands:

\begin{enumerate}
	\item get 	 - Print the value of a variable
		Arguments: [VARIABLE NAME]
	\item set 	 - Modify the value of a variable
		Arguments: [VARIABLE NAME] [NEW VALUE]
	\item fit 	 - Fit the contents of an ASCII data file
		Arguments: [FIT TYPE]
	\item sim 	 - Simulate a FID from scratch
		Arguments: [NUMBER OF STEPS]
	\item help	 - Print this help
		Arguments: <none>
	\item exit	 - Quits the program
	  Arguments: <none>
\end{enumerate}

\noindent
The \textit{get} and \textit{set} commands are meant to respectively print and modify the value of the variable whose name is passed as an argument. Here is a list of the valid variable names:

\begin{itemize}
\item \textbf{b\_{min}, b\_0, b\_{max}} - Boundaries and central value for coupling constant b (kHz)
\item \textbf{D\_min, D\_0, D\_max} - Boundaries and central value for diffusivity D ($\mathrm{ms}^{-1}$)
\item \textbf{X\_min, X\_0, X\_max} - Boundaries and central value for X = D/b (values change accordingly to follow b and D)
\item \textbf{t\_min, t\_0, t\_max} - Boundaries and central value for time (variable unit). Used as limits in fits and simulations. t\_max = -1 means that there is no upper limit
\item \textbf{poles} - Number of poles for FID truncation
\item \textbf{iters} - Number of simplex iterations for fitting
\item \textbf{thr} - Threshold for calculating fitting convergence
\item \textbf{t\_col, data\_col} - Indices for columns of time and data in input file, respectively
\item \textbf{t\_unit} - Time unit in input file (seconds, standard value = 1E-3 s)
\item \textbf{norm} - Normalization factor for data in input file
\item \textbf{skip\_l} - Lines to skip at the beginning of the input file
\item \textbf{start\_p, end\_p} - Starting and ending data points in input file. end\_p = -1 means that there is no upper limit
\item \textbf{skip\_p} - Point skipping step in input file
\item \textbf{input\_file} - Input file name
\item \textbf{output\_file} - Output file name
\end{itemize}

The meaning of most of the variables is self-explaining. The parameter $X=D/b$ is introduced because it is easier to grasp its value by eye and is best fitted than $D$ - in general, $X=0.01$ means an almost perfectly rigid FID, while $X=10$ is a completely mobile one. The central values, \textit{b\_0, D\_0} and \textit{X\_0} are used as fixed values in simulations, while in fittings they matter only when a variable is kept constant. The boundaries apply to fittings and are not relevant to simulations. Of course, since the $D$ and $X$ variables are interdependent, any change applied to one will reflect on the other. The $t$ variables are slightly different: in fittings, the boundaries represent the time limits of the data file on which the residual sum of squares is calculated, in simulations the time interval to simulate. If there is no upper limit, simulations will use a default calculated value.\newline
Number of poles, of simplex iterations, and threshold for convergence (basically the criterion of acceptance of a local minimum in the Nelder-Mead algorithm) have optimal default values that usually do not need to be changed. For the fitting, the raw data must be contained in an ASCII data file. The \textit{t\_col} and \textit{data\_col} are indices of the columns containing respectively the time and the FID data; it is possible to have $data\_col < t\_col$, but they must not be equal. It is possible as well to configure a number of lines to skip, for example to remove a textual header from the data file (\textit{skip\_l}), or to fit only on one point each \textit{skip\_p} points - this is useful to speed up the fitting if the raw file has a high time resolution. It is possible to scale both time (\textit{t\_unit}) and data (\textit{norm}) by a constant factor; it must be remembered that the data has to be normalized in a way that it goes to 1 at $t=0$ for the fitting to work. Finally, the paths of the file
to read (\textit{input\_file}) and to write (\textit{output\_file}) can be inserted. When inserting the names, no apices or quotation marks should be used. At the beginning of each file name the program will add either \textit{fitted\_} or \textit{simulated\_} in order to make the files recognizable and prevent accidental overwriting.\newline
For some of these variables there are values that are not acceptable (for example, $start\_p < 0$) and values that conflict with other variables (for example, $D\_min > D\_max$). When a not acceptable value is inserted with the function \textit{set}, an error message is printed and the value of the variable is not changed. Every time a variable is changed with \textit{set} the new value is printed immediately to confirm the effect of the command.\newline
The command \textit{fit} accepts one argument of type \textit{FIT TYPE}. This is simply a string which can assume three values: \textit{b}, \textit{D} and \textit{bD}. Its purpose is to indicate which fitting parameters must be found. In this way, for example, inserting:

> fit bD

\noindent
will run a two dimensional fitting on both $b$ and $D$, while the input:

> fit D

\noindent
will only fit $D$ while keeping $b=b\_0$.\newline
The command \textit{sim} requires for an argument only the number of steps for the simulation. Remember that, however, the \textit{length}, in time units, of the simulation is controlled by \textit{t\_min} and \textit{t\_max}: the number of steps will only affect the resolution with which the FID is simulated on this interval.\newline
If one needs to fit many data files with similar structure in one go, then it becomes convenient to make use of the possibility to control the program via input file. This is done simply by running the program with the files to be fitted as arguments into the system console:

user:~\$ eddie file\_to\_fit\_1.dat file\_to\_fit\_2.dat ...

\noindent
In this case, there must be, in the same folder of the program, a file \textit{parameters.txt}, which will contain the instructions for the fitting. These instructions will simply be a sequence of commands as the ones described earlier: the program will simply run all these instructions in sequence in its console before performing the fitting. The program is \textit{not} able to simulate using an input file, and it is not necessary to insert a \textit{fit} line at the end of the input file. The only difference between using the program with the console or with an input file is that in the latter case, an instruction to \textit{set} either $b\_0$, $D\_0$ or $X\_0$ will fix the set variable for the fitting. In other words, if one wants the fitting to find both $b$ and $D$, there must be no \textit{set} instructions for these three variables; on the other hand, setting, for example, $b\_0$, will result in a fitting running only on the parameter $D$ while keeping $b=b\_0$.

Finally, the structure of the output files is rather straightforward. In both a simulation and a fitting, the file begins with a two lines header. The first line contains the values of $b$ and $D$ used (either the fixed ones in a simulation, or the fitted ones in a fitting) and the second is just an indication of what the various columns contain:

b = 386.586 kHz	 D = 1737.96 kHz

Time (ms)	Data	Fitted

In this case, the example was taken from a fitting file, and there are three columns: time, data (the original fitted data) and fitting (the best fitting FID found by the program). In the case of a simulation, there would be only two columns, time and FID. The fitting file is produced always in the same folder as the file containing the original data. When fitting from an input file, the program also produces a further output file, in its own working directory, called \textit{results.txt}. This file will contain three columns with, respectively, the names of the fitted files and the found values for $b$ and $D$.


\clearpage


\begin{thebibliography}{10}

\bibitem{degenn}
P.~de~Gennes, {\em Scaling concepts in polymer physics}.
\newblock Cornell University Press, 1979.

\bibitem{doi_edw}
{M. Doi} and {S.F. Edwards}, {\em The theory of polymer dynamics}.
\newblock Clarendon Press, 1994.

\bibitem{rel_proc_superc}
{W. G\"{o}tze} and {L. Sj\"{o}gren}, ``Relaxation processes in supercooled
  liquids,'' {\em Rep. Prog. Phys.}, vol.~55, pp.~241--376, 1992.

\bibitem{gl_diel}
{A. Kudlik}, {S. Benkhof}, {T. Blochowicz}, {C. Tschirwitz}, and {E.
  R\"{o}ssler}, ``The dielectric response of simple organic glass formers,''
  {\em Jour. Mol. Struct.}, vol.~479, pp.~201--218, 1999.

\bibitem{PB_decoup}
{D. Richter}, {R. Zorn}, {B. Farago}, {B. Frick}, and {L.J. Fetters},
  ``Decoupling of time scales of motion in polybutadiene close to the glass
  transition,'' {\em Phys. Rev. Lett.}, vol.~68, pp.~71--74, 1992.

\bibitem{PB_diel_data}
{R.D. Deegan} and {S.R. Nagel}, ``Dielectric susceptibility measurements of the
  primary and secondary relaxation in polybutadiene,'' {\em Phys. Rev. B},
  vol.~52, pp.~5653--5656, 1995.

\bibitem{PB_liq_melt}
{S. Kariyo}, {C. Gainaru}, {H. Schick}, {A. Brodin}, {V.N. Novikov}, and {E.A.
  R\"{o}ssler}, ``From a simple liquid to a polymer melt: Nmr relaxometry study
  of polybutadiene,'' {\em Phys. Rev. Lett.}, vol.~97, p.~207803, 2006.

\bibitem{PB_liq_melt_2}
{S. Kariyo}, {A. Brodin}, {C. Gainaru}, {A. Hermann}, {H. Schick}, {V.N.
  Novikov}, and {E.A. R\"{o}ssler}, ``From simple liquid to a polymer melt.
  glassy and polymer dynamics studied by fast field cycling nmr relaxometry:
  Low and high molecular weight limit,'' {\em Macromolecules}, vol.~41,
  pp.~5313--5321, 2008.

\bibitem{pier_stu_FID}
{S. Sturniolo} and {M. Pieruccini}, ``An exact analytical solution for the
  evolution of a dipole-dipole interacting system under spherical diffusion in
  a magnetic field,'' {\em J. Mag. Res.}, vol.~223, pp.~138--147, 2012.

\bibitem{MSE_stur}
{S. Sturniolo} and {K. Saalw\"{a}chter}, ``Breakdown in the efficiency factor
  of the mixed magic sandwich echo: A novel nmr probe for slow motions,'' {\em
  Chemical Physics Letters}, vol.~516, pp.~106--110, 2011.

\bibitem{NBR_publ}
{S. Sturniolo}, {M. Pieruccini}, {M. Corti}, and {A. Rigamonti}, ``Probing
  $\alpha$-relaxation with nuclear magnetic resonance echo decay and
  relaxation: a study on nitrile butadiene rubber,'' {\em Solid State NMR},
  vol.~51--52, pp.~16--24, 2013.

\bibitem{demco}
R.~Fechete, D.~Demco, and B.~Bl\"{u}mich, ``Chain orientation and slow dynamics
  in elastomers by mixed-magic-hahn echo decyas,'' {\em The Journal of Chemical
  Physics}, vol.~118, pp.~2411--2421, 2003.

\bibitem{mse_seminal}
{W-K. Rhim}, {A. Pines}, and {J. S. Waugh}, ``Time-reversal experiments in
  dipolar-coupled spin systems,'' {\em Physical Review B}, vol.~3, no.~3,
  pp.~684--696, 1971.

\bibitem{brer}
M.~Brereton, ``An exact expression for the transverse nuclear magnetic
  resonance relaxation of a dynamic scale invariant polymer chain governed by a
  single relaxation time,'' {\em J. Chem. Phys.}, vol.~94, no.~3,
  pp.~2136--2142, 1991.

\bibitem{rel_proc_geom}
{U. Tracht}, {A. Heuer}, and {H.W. Spiess}, ``Geometry of reorientational
  dynamics in supercooled poly(vinyl acetate) studied by $^{13}${C}
  two-dimensional nuclear magnetic resonance echo experiments,'' {\em J. Chem.
  Phys.}, vol.~111, pp.~3720--3727, 1999.

\bibitem{papon}
{A. Papon}, K. Saalw\"{a}chter, K. Sch\"{a}ler, L. Guy, F. Lequeux and
H. Montes, ``Low-field nmr investigations of nanocomposites:
  Polymer dynamics and network effects,'' {\em Macromolecules}, vol.~44,
  pp.~913--922, 2011.

\bibitem{A_W}
P.~Anderson and P.~Weiss, ``Exchange narrowing in paramagnetic resonance,''
  {\em Rev. of Modern Physics}, vol.~25, pp.~269--276, 1953.

\bibitem{logMOM}
R.~Zurn, ``Logarithmic moments of relaxation time distributions,'' {\em J.
  Chem. Phys.}, vol.~116, pp.~3204--3209, 2002.

\bibitem{colmenero}
A. Arbe, D. Richter, J. Colmenero and B. Farago,
``Merging of the $\alpha$ and $\beta$ relaxations in polybutadiene: A neutron spin echo and dielectric study'' {\em Phys. Rev. E}, vol.~54, 3853--3869, 1996.

\bibitem{phillips}
J. C. Phillips,
``Stretched exponential relaxation in molecular and electronic glasses'' {\em Rep. Prog. Phys.}, vol.~59, 1133–-1207, 1996

\bibitem{colmeneroHN}
F. Alvarez, A. Alegr$\acute{\textrm{\i}}$a and J. Colmenero,
``Relationship between the time-domain Kohlrausch-William-Watts and frequency domain Havriliak-Negami relaxation functions'' {\em Phys. Rev. B}, vol.~44(14), 7306--7312, 1991.

\bibitem{correlations}
S.W. Morgan, V. Oganesyan and G.S. Boutis,
``Multispin correlations and pseudothermalization of the transient density matrix in solid-state NMR: Free induction decay and magic echo'' {\em Phys. Rev. B}, vol.~86, 214410, 2012.

\bibitem{hvks}
H.~Vestermark, ``http://www.hvks.com/index.html.''

\bibitem{GBNM}
{M.A. Luersen}, {R. Le Riche}, and {F. Guyon}, ``A constrained, globalized, and
  bounded Nelder-Mead method for engineering optimization,'' {\em Struct.
  Multidisc. Optim.}, vol.~27, pp.~43--54, 2004.

\end{thebibliography}
\end{document}